# Monitoring of SF$_6$ Degradation in GIS Using Frequency Response Analysis

Mehdi Babaei and Ahmed Abu-Siada
Curtin University, Perth, Western Australia

*Abstract*—Frequency Response Analysis (FRA) is generally used to detect any changes in the active part of the test object. Several studies have analyzed the effect of different mechanical changes including deformations and oil insulation decays on frequency response of transformers as a test object, however noresearch has ever studied the effect of gas degradation on frequency signature of a gas insulated substation (GIS). This paper investigates the possibility of using frequency response analysis to assess the quality of insulation medium in gas insulated substation (GIS). For this purpose, a single phase GIS filled with SF$_6$ between the main conductor and the enclosure is simulated in Maxwell 3D using finite element method to obtain the electrical parameters of the substation in real mode operation. Measurement of frequency response was carried out in GIS simulations with different insulation qualities from healthy to weak condition. The results show that the frequency response of GIS can disclose any minor deterioration in gas quality of the switchgear.

## I. INTRODUCTION

SULPHUR hexafluoride (SF$_6$) is a non-toxic, inert, insulating and cooling gas of high dielectric strength and thermal stability which is used in gas insulated switchgears as insulation medium [1]. GIS insulation aging can reduce the dielectric strength of SF$_6$ due to frequent operation of disconnecting switches and circuit breakers [2, 3] which will result in some catastrophic failures if a reliable monitoring system is not applied to assess the quality of SF$_6$ regularly. Frequency response analysis (FRA) is one of the methods used in other high voltage applications such as power transformer to detect any mechanical faults and geometrical changes of internal components including coils and winding [4]. Some examples of conditions that FRA can be used to assess are damages following short-circuit or any faults during transportation, andfollowing a seismic event.

The main concept used in FRA is that any mechanical changes in mechanical features of the test object results in some change in electrical parameters of the equivalent model of the object then results in some changes to the frequency response of the object. Existing methods to determine SF$_6$ degradation are based on measurements by several sensors connected to GIS, but hybrid systems compromising measurement and simulation methods have drawn considerable attention for the purpose of asset management in electrical plants [5]. The detection of damage using FRA is most effective when frequency response measurementdata is available from the object when it is in a known good condition which can be considered as a FRA signature of the object and can be used for assessing the condition of the object during its continuous operation [6]. Regarding the importance of gas insulation monitoring in GIS, this method can be applied in GIS to detect any probable dielectric deterioration in the substation which can FRA signature of GIS disclose.

To make a frequency response measurement, a low voltage signal is applied to one terminal of the test object with respect to the tank. The voltage measured at this input terminal is used as the reference signal and a second voltage signal (the response signal) is measured at a second terminal with reference to the tank. The frequency response amplitude is the scalar ratio between the response signal ($V_{out}$) and the reference voltage ($V_{in}$) (presented in dB) as a function of the frequency. The phase of the frequency response is the phase difference between $V_{in}$ and $V_{out}$ (presented in degrees). The measured FRA signature could be in the form of impedance instead of transfer function as well [7].

## II. SF$_6$ DEGRADATION

High dielectric strength of SF$_6$is due to high electronegativity of fluorine, it means that the life span of the free electrons remains very low and with the SF$_6$molecules they form heavy ions with low mobility. Then the probability of dielectric failure by a snowballing effect and electron removal from the gas under electrical field is delayed [8]. Also the thermal behavior of SF$_6$is very stable due to the symmetrical shape of its molecules with one sulfur atom at the center and six fluorine atoms at the corners of the molecule. Following a breakdown, SF$_6$ regenerates itself. Its original strength is spontaneously restored and, in most cases, is even slightly enhanced. Due to the very low adiabatic coefficient of SF$_6$, the pressure rise as aresult of thermal expansion following dielectricbreakdowns is less than that with othergases and very considerably less than is thecase with liquid dielectrics[9].These high quality feature of SF$_6$ made it particularly suitable for application in power circuit breakers and high-voltage switchgears as well as in high voltage cables and transformers [8].

During the arcing time in $SF_6$ with very high temperature about 20,000°K, the gas breaks down and although $SF_6$ is chemically stable material, it decomposes into low number fluorine molecules such as $SF_2$ which are extremely toxic, corrosive materials. This created impurities can also deteriorate the insulating performance of $SF_6$ by reducing the dielectric strength of $SF_6$[10]. Since gas insulated substations have been extensively used in power networks within recent few decades with $SF_6$ as insulating medium in an environment with large number of switching operation and frequent strikes due to their higher reliability and safety [2], special attention must be given to the condition monitoring of $SF_6$ in gas insulated substations.

Moisture ingress in GIS may have deteriorating effect on the insulation medium. Very high levels of humidity increase the possibility that water molecules will condense into the liquid phase, adversely affecting the dielectric withstand strength of GIS [11] and significantly increasing the possibility of flashover. Therefore, humidity in GIS should be maintained at a level such that the humidity does not condense in the form of liquid water over the entire range of the expected operating temperatures and gas pressures. Also the formation of corrosive and toxic decomposition by-products occurs in reactions between moisture and dissociated $SF_6$ found in the high energy arcs during normal switching operations. Higher concentrations of humidity can result in higher concentrations of decomposition by products [B3]. These by-products cause corrosion and may degrade insulators and other internalcomponents within the GIS, and pose a safety hazard.

The dielectric property of insulation system is characterizedbased on two fundamental factors; permittivity and conductivity. While the permittivity affects insulation system behaviour during electrical transient conditions, electric conductivity plays an essential role to specify the dielectric strength of insulating system [12]. $SF_6$ degradation is usually accompanied with change in both insulation permittivity and conductivity which results in change in the capacitance between GIS main conductor and the enclosure. This change in capacitance value can be described by [13]:

$$\epsilon^* = \frac{\epsilon}{\epsilon_0} = \epsilon_r - j\epsilon'' = \epsilon_r - j\frac{\sigma}{\omega} = \frac{C(\omega)}{C_0} \quad (1)$$

where, $\epsilon$, $\epsilon^*$, $\epsilon_0$ and $\epsilon_r$ refer to the total, complex, vacuum and relative permeability, respectively. $\sigma$ is the electrical conductivity (S.m-1) and $\omega$ is the frequency (rad.s-1). Eq. (1) clarifies that GIS capacitance is significantly dependent on dielectric permittivity and conductivity. Any change in these factors will be reflected as a kind of change in capacitance.Table 1 shows the electrical characteristics of $SF_6$.

TABLE 1. CHARACTERISTICS OF SF6

| Parameter | Description |
| --- | --- |
| Dielectric breakdown | 89 V/mPa |
| Relative permittivity | 1.00204 |
| Electrical conductivity | 1e-16 Siemens/m |

## III. GIS MODELLING USING FINITE ELEMENT METHOD

For the purpose of studying the effect of $SF_6$ aging or degradation on frequency response signature of GIS, $SF_6$ permittivity and conductivity are changed to particular levels. These minor changes in insulation characteristics of $SF_6$, reflect as changes in the series capacitance of GIS conductor which can appear in changes in distributed parameter model of GIS.In this paper, the feasibility of utilizing FRA in recognizing SF6 degradation is investigated through the simulation of a real 400kV gas insulated substation located in the north of Iran. The single line diagram and arrangement of equipment are obtained from [14, 15]. Arrangement of GIS equipment and gas compartments are shown in Fig.1. It is shown in Fig.1 that each transformer feeder is assembly of several high voltage equipment such as circuit breaker, disconnecting switch and instrument transformers connected in double busbar arrangement with bypass disconnecting switch. Each equipment is located in its own gas compartment enclosed by aluminum enclosure. The space between GIS conductor and the enclosure contains $SF_6$ as insulating medium.

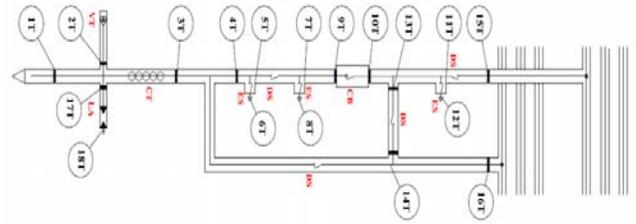

Fig. 1. Gas compartments of GIS equipment

According to Fig. 2, all GIS sections are modelled as distributed parameter line with R, L, and C parameters calculated by finite element method. It means that all GIS compartments are simulated in Maxwell 3D software in three dimensional real model as shown in Fig. 3 to Fig. 5 and the capacitance and inductance of each compartment are computed using electrostatic and magnetostatic solvers of the software respectively.

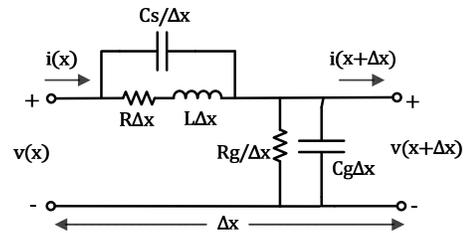

Fig. 2. Distributed parametr model of GIS sections

All solids modelled in the software are meshed by adaptive meshing operation and inorderto obtainthesetofalgebraicequationstobesolved,thegeometryo ftheproblemisdiscretizedautomaticallyintosmallelements which are called tetrahedral due to their shape and are produced by meshing operation. Capacitive elements are calculated by running the electrostatic solver of Maxwell 3D where voltage V is applied on GIS conductor, while the voltage level is maintained at a level of zero on the aluminium enclosure. The energy stored in electrostatic field ($W_{ij}$) between the two elements can be calculated as follows [16]:

$$W_{ij} = 0.5 \int_{Vol} D_i \times E_j \, dVol \quad (2)$$

Where $W_{ij}$ is the electrical field energy between elements $i$ and $j$, $D_i$ is the electrical flux density of element $i$ and $E_j$ is the electrical field intensity of element $j$ and $Vol$ is the volume of simulated object. The capacitance $C_{ij}$ between two elements $i$ and $j$ can then be calculated as:

$$C_{ij} = 2 \times \frac{W_{ij}}{V^2} \quad (3)$$

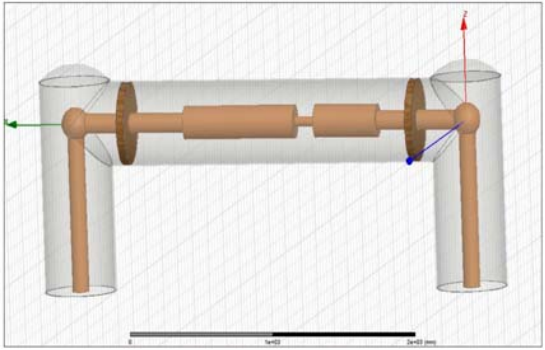

Fig. 3. Closed circuit breaker model in Maxwell 3d

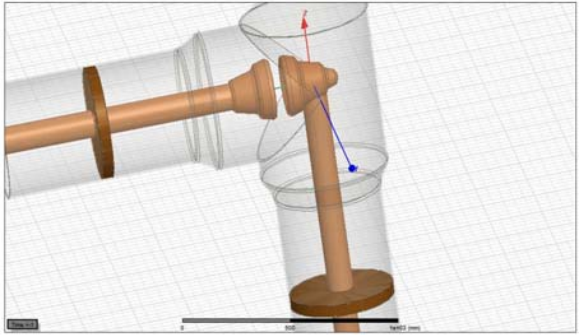

Fig. 4. Disconencting switch model

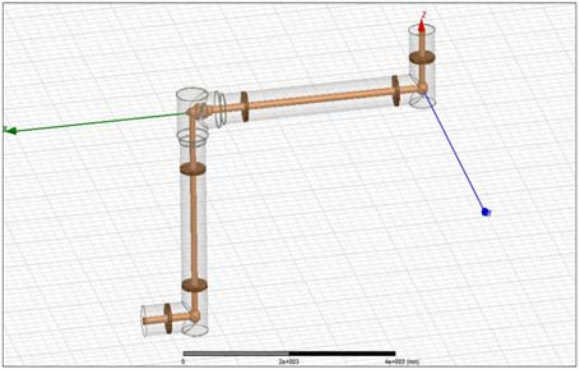

Fig. 5. 3D model of GIS connections

Inductive components are calculated based on the average of the magnetic field energy ($W_{ij}$) and the corresponding peak current passing through the conductor ($I_P$) as follows[16]:

$$W_{ij} = 0.5 \int_{Vol} B \times H \, dVol \quad (4)$$

$$L_{ij} = 2 \times \frac{W_{ij}}{I^2} \quad (5)$$

where $B$ is the magnetic field density, $H$ is the magnetic field intensity, and $Vol$ is the conductor volume. Resistive components are calculated based on power losses ($P_{loss}$) depending on conductor conductivity ($\sigma$) and current density ($J$) as given in (5) and (6) below [12]:

$$P_{loss} = \left(\frac{1}{2\sigma}\right) \int_{Vol} J.J \, dVol \quad (6)$$

$$R = P_{loss}/I^2 \quad (7)$$

Table 2 in lists the calculated GIS equivalent circuit parameters.

TABLE 2. GIS PARAMETERS

| Parameter | Description |
|---|---|
| External radius of GIS conductor | 60 mm |
| Internal radius of GIS conductor | 42.5 mm |
| External radius of enclosure | 254 mm |
| Internal radius of enclosure | 246.1 mm |

## IV. IMPACT OF SF$_6$ DEGRADATION ON GIS FRA SIGNATURE

In this section, three different levels are defined for SF$_6$ degradation as presented in Table 3: slight, moderate and significant variation in gas dielectric characteristics based on 5%, 10% and 15% change in permittivity of SF$_6$ respectively. Then the impact of gas degradation on FRA signature of GIS is studies by applying frequency response analysis on GIS switchgear. FRA signature of GIS is presented in Fig. 6 by applying the FRA measurement on healthy insulation system and is considered as a base case for being compared with the frequency response of degraded SF$_6$. The signature is obtained by sweeping an AC low voltage source and variable frequency (up to 10 MHz) on the terminal of SF$_6$/Air bushing in line feeder or cable termination in transformer feeder and calculating the input current. The FRA signature is plotted in the form of GIS input impedance in dB. ($Z_{in}=log_{10}(V_{in}/I_{in})$) for varying frequency. Then the FRA of GIS is obtained by repeating the described procedure in each defined case of degradationlevel.

TABLE 3. SF$_6$ DEGRADATION LEVEL

| Level | Base | slight | moderate | Significant |
|---|---|---|---|---|
| Relative permittivity | 1 | 1.1 | 1.2 | 1.3 |

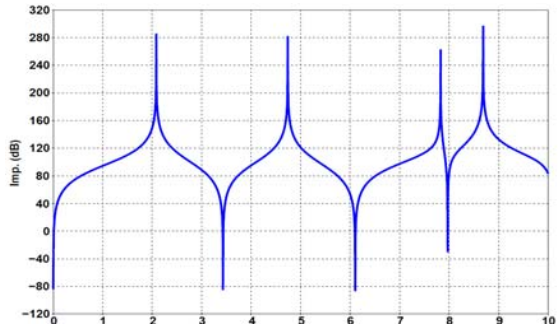

Fig. 6. FRA signature of GIS as base case

As shown in Fig. 6, there are several resonance and anti-resonance appearing in FRA signature of GIS and at higher frequencies the resonance frequencies are much less significant due to dominance of capacitive characteristics of GIS components in comparison with inductive characteristics of GIS conductor. Fig. 7 to Fig. 9 show the FRA of GIS in other health conditions of $SF_6$ and compares the FRA with the base case. The impact of $SF_6$ degradation is reflected as slight shift in FRA signature toward low frequencies with respect to the FRA of the base case and the severity of this shift is dependent on the severity of $SF_6$ degradation.

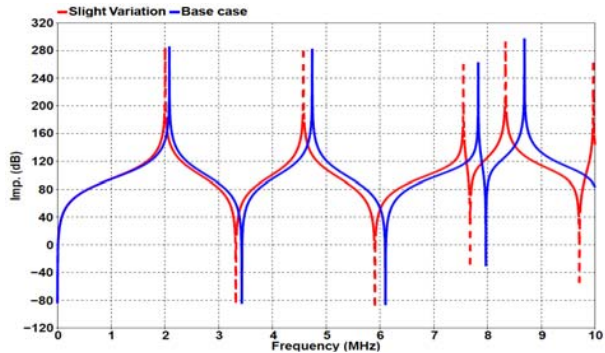

Fig. 7. Slight $SF_6$ degradation on FRA signature

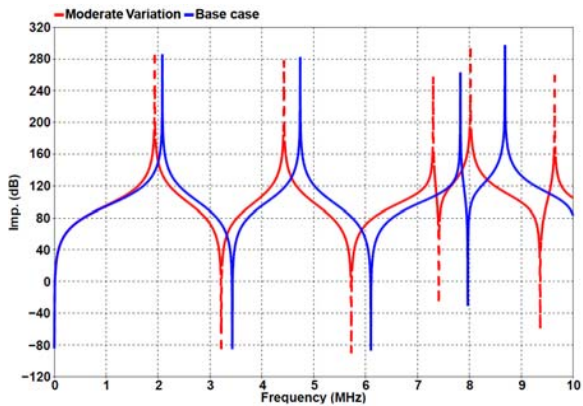

Fig. 8. Moderate $SF_6$ degradation on FRA signature

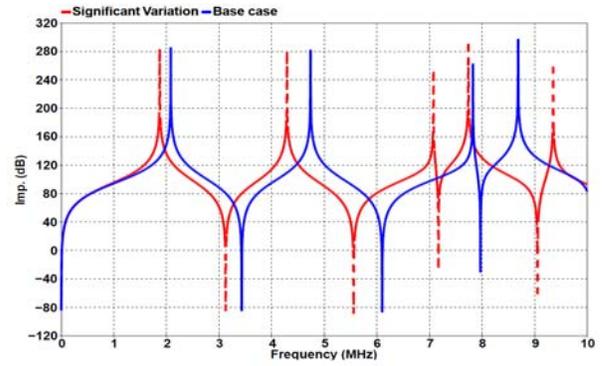

Fig. 9. Signifiant $SF_6$ degradation on FRA signature

The above results show that FRA has a potential to detect SF6 aging / degradation. The effect of oil degradation is noticeable for frequency ranges above 1MHz where resonance frequencies tend to shift to the left. The impact is more pronounced with the increase of $SF_6$ degradation level.

## V. CONCLUSION

In this study, the concept of using the frequency response of an object for the purpose of detecting the changes in electrical characteristics of that object was applied to gas insulated substation to detect the amount of degradation level of $SF_6$. Gas insulated substation was simulated in EMTP-RV by distributed parameter model of equipment. The parameter were calculated by modelling in three dimensional real geometry and running finite element analysis. Simulation results show that $SF_6$ aging / degradation introduces a reduction to the resonance and anti-resonance frequencies in the high frequency range. Also, the magnitude of the FRA signature is slightly decreased within the mid frequency range. Furthermore, this method has advantage over other existing methods of using gas system sensors on local control cabinets of GIS regarding the higher reliability of the proposed FRA signature method.